# Accommodation mechanisms in strain-transformable titanium alloys

Y. Danard[a], G. Martin[b], L. Lilensten[a*], F. Sun[a], A. Seret[c], R. Poulain[a], S. Mantri[d], R. Guillou[f], D. Thiaudière[e], I. Freiherr von Thüngen[g], D. Galy[g], M. Piellard[g], N. Bozzolo[c], R. Banerjee[d], F. Prima[a]

[a] PSL Research University, Chimie ParisTech, Institut de Recherche de Chimie Paris, CNRS UMR 8247, 75005 PARIS, France

[b] Université Grenoble Alpes, CNRS UMR 5266, Grenoble INP, Laboratoire SIMaP, Grenoble, 38000, France

[c] MINES ParisTech, PSL - Research University, CEMEF - Centre de mise en forme des matériaux, CNRS UMR 7635, 06904 Sophia Antipolis Cedex, France

[d] Department of Materials Science and Engineering, University of North Texas, Denton, TX-76207, USA

[e] DEN-Service de Recherches Métallurgiques Appliquées (SRMA), CEA, Université Paris-Saclay, F-91191, Gif-sur-Yvette, France

[f] Synchrotron SOLEIL, l'orme des merisiers, Saint-Aubin BP48, F-91192 Gif-sur-Yvette, France

[g] SAFRAN TECH, Rue des Jeunes Bois, Châteaufort - CS 80112 - 78772 Magny-Les-Hameaux, France

* corresponding author: Lola LILENSTEN

lola.lilensten@chimieparistech.psl.eu

IRCP - Chimie ParisTech, 11, Rue Pierre et Marie Curie 75005 PARIS

**Abstract**

A new β-metastable Ti-alloy is designed with the aim to obtain a TWIP alloy but positioned at the limit between the TRIP/TWIP and the TWIP dominated regime. The designed alloy exhibits a large ductility combined with an elevated and stable work-hardening rate. Deformation occurring by formation and multiplication of {332}<113> twins is evidenced and followed by in-situ electron microscopy, and no primary stress induced martensite is observed. Since microstructural investigations of the deformation mechanisms show a highly heterogeneous deformation, the reason of the large ductility is then investigated. The spatial strain distribution is characterized by micro-scale digital image correlation, and the regions highly deformed are found to stand at the crossover between twins, or at the intersection between deformation twins and grain boundaries. Detailed electron back-scattered imaging in such regions of interest finally allowed to evidence the formation of thin needles of stress induced martensite. The latter is thus interpreted as an accommodation mechanism, relaxing the local high strain fields, which ensures a large and stable plastic deformation of this newly designed Ti-alloy.

**Highlights**

- A novel TWIP Ti-alloy is designed
- Heterogeneous localization of the deformation is evidenced.
- Highly-strained areas are investigated
- Micron-size martensite needles form in these areas
- Martensite accommodates local strains and prevents early fracture

1. Introduction

In recent years, strain-transformable titanium alloys, with improved ductility and very high strain-hardening rate have been the subject of extensive investigations [1–5]. The elementary mechanisms such as TRIP (transformation induced plasticity) and TWIP (twinning induced plasticity) have been more particularly investigated, as a function of both alloy chemistry and strengthening approaches (precipitation mechanisms), see e.g. [6–9]. However, the underlying physical mechanisms governing the mechanical properties in those metastable alloys is still debated in the literature: if it is now well-established that the high strain-hardening rate is closely related to the dynamic formation of a dense network of mostly {332}<113> twins, leading to a strong dynamic Hall-Petch effect, by drastic reduction in the mean free path of mobile dislocations [1,2,10,11], yet, the microstructural evolution, upon loading, appears to be extremely rapid, complex and critically dependent on the investigated TRIP/TWIP Ti-alloy [11,12]. As a matter of fact, the macroscopic strain-hardening depends on a set of underlying mechanisms, which are presently not completely understood, justifying the need for further research.

Another unclear aspect of deformation in TRIP/TWIP Ti-alloys is related to the comparative observations of deformation behavior at mesoscopic/macroscopic scales, showing an obvious paradox between the extremely large macroscopic ductility and the strong strain localization at the mesoscopic and microscopic scale: (1) from one grain to its neighbors [3], (2) at the twin/matrix interfaces, and, (3) at the intersection of multiple twin variants. Fundamentally, local deformation is extremely heterogeneous in strain-transformable Ti-alloys, and the grades exhibiting a TRIP/TWIP effect display a remarkable tolerance with respect to strong internal stress concentrations, suggesting that accommodation mechanisms are triggered during the deformation to reach extremely large ductility. It has been shown that deformation mechanisms are of two kinds, depending on whether they are activated to accommodate the external loading, or the internal stresses arising from the intrinsic heterogeneous nature of the deformation mechanisms at the mesoscopic scale. The latter are of several types, such as secondary twinning systems, martensitic transformation, stress-induced omega phase or geometrically necessary dislocations [2,3,7,8,13]. In particular, the role of stress induced $\alpha''$ martensite has recently been linked to the accommodation of the mechanical contrast between the primary $\alpha$-phase precipitates and the $\beta$-matrix, in a dual-phase Ti-10V-2Fe-3Al (Ti-10-2-3) [7].

This raises the question of a potential trade-off that has to be achieved between complementary hardening and relaxation mechanisms, to reach an optimized combination of high strength, strain-hardening, and extended range of stable plastic deformation, leading to substantial ductility.

The present paper provides a new insight into the role of stress-induced martensite as a possible accommodation mechanism in TRIP/TWIP Ti-alloys. A new Ti-alloy showing an intense TWIP effect to accommodate external loading, but with no primary $\alpha''$ martensite precipitation, is deliberately designed, with composition Ti-7Cr-1Al-1Fe (wt%) (named TCAF hereafter). This alloy displays a remarkable combination of ductility and work-hardening. The deformation mechanisms are investigated to rationalize the ductility, within the proposed frame of "accommodation mechanisms".

2. **Materials and Methods**

*2.1 Materials and processing conditions*

A Ti-7Cr-1Al-1Fe (wt%) (TCAF) button (200g) was prepared using a tungsten arc-melting furnace under high-purity Ar atmosphere. After hot deformation in $\alpha/\beta$ domain, oxygen content is measured to be 1072 ppm by fusion under inert gaz (ASTM E1409 test method). Then, the ingot was solution-treated at 1173 K for 1.8 ks under air and water quenched to reach a full $\beta$- phase microstructure. After careful surface polishing to remove the oxygen enriched layer due to the high-temperature thermal treatment, cold rolling was performed to manufacture sheets with a thickness of 0.65 mm (thickness reduction of 85%). Cold-rolled sheets were then solution-treated in the β domain, at 1073 K for 900 s, under high purity Ar atmosphere and immediately water quenched. The produced specimens exhibited a fully β recrystallized microstructure with an average grain size of 100 µm (see EBSD orientation map of Figure S1), defined as the mean circle diameter of the detected grains with a misorientation angle threshold of 10° [14].

*2.2 Microstructural and mechanical characterizations*

Tensile specimens were extracted from the metal-sheets, by wire EDM, with gauge dimensions of 50 x 4 x 0.65 mm$^3$. Uniaxial tensile tests (monotonic and cyclic tests) were carried out at room temperature along the rolling direction, using an INSTRON5966 machine with a 10 kN load cell and an external extensometer with a gauge length of 10 mm, at a strain rate of 1.7 x 10$^{-3}$ s$^{-1}$. *In-situ* tensile tests for electron backscatter diffraction (EBSD) were performed using a Proxima 100-Micromecha machine on plate-shaped samples with gauge dimensions 35 x 2 x 0.65 mm$^3$, whose surface was prepared by electropolishing. EBSD scans were collected using a ZEISS LEO-1530 field emission gun scanning electron microscope (SEM) operating at 20 kV equipped with a NORDIFF system. A step size of 2 µm was employed to characterize the microstructure in the undeformed conditions, and of 0.8 µm after strain increments (3% and 4% of macroscopic tensile strain). Microscale strain heterogeneities were measured during straining by *in-situ* tensile tests conducted in in a ZEISS GeminiSEM500 using a Deben MT2000 EW tensile machine, at a strain rate of 3 x 10$^{-4}$ s$^{-1}$. The micro tensile specimens with gauge dimensions of 30 x 1.5 x 0.65 mm$^3$ were machined by wire EDM and then mirror-polished with SiC papers and the colloidal silica (particle size 0.04µm) solution. To be able to perform microscale digital image correlation (DIC), a speckle was deposited prior to mechanical testing on the surface using the colloidal silica solution as indicated in [15]. Successive high-definition images (8192 x 6459 pixels) of the same region of interest were acquired after different strain increments (undeformed conditions, 0.61%, 1.14% and 1.98% macroscopic tensile strain) using secondary electron (SE) contrast with an accelerating voltage of 15 kV. The macroscopic tensile strain indicated above are average values of the local tensile component $\varepsilon_{11}$ over the studied area calculated from the 90 000 points where the local strains were computed. High-definition imaging (pixel size ~22 nm) was required to achieve a good description of the local contrast combined with a relatively large field of view to ensure the statistical consistency of the local strain mapping. The comparison between undeformed and deformed states allows the local strains to be computed. DIC was performed using the CMV software [16,17]. For a more detailed description of the DIC routine applied here, the reader can refer to Martin *et al.* [18,19] and Lechartier *et al.* [20]. Finally, 2% and 5% deformed TCAF samples were also electropolished to record an EBSD map on the deformed specimen, using a Bruker Crystalign system mounted onto a Zeiss SUPRA40 field emission gun scanning electron microscope operating at 20 kV, with a step size of 94 nm. The latter EBSD map has been analyzed to quantify the spatial distribution of geometrically necessary dislocations (GND) using the method suggested by Seret et *al*. [21].

### 3. Results and discussion

*3.1 Design strategy*

In this paper, a new strain-transformable alloy, Ti-7Cr-1Al-1Fe (in wt%) has been designed, using the d-electron method, commonly called the Bo-Md method [22]. The composition was initially designed to be on the Md=RT line, corresponding to the limit of the stress-transformation range for α" stress-induced precipitation, hence targeting alloys with low TRIP effect. However, unlike previous studies [1–4,23], this method has been combined with an additional design approach, aiming at limiting the direct TRIP effect and to correlatively increase the yield strength. This additional design strategy is based on the $[Fe]_{eq}$ empirical parameter recently proposed by Bignon et *al.* [24]. This empirical parameter has been proven to be efficient in predicting the presence of TRIP effect as a primary deformation mechanism in strain-transformable Ti-alloy. According to Bignon et *al.* [24], the $[Fe]_{eq}$ parameter depends on the alloy's composition and is defined according to the following equation:

$$[Fe]_{eq} = 3.5 \, x \, (\frac{[Fe]}{3.5} + \frac{[Cr]}{9} + \frac{[Mo]}{14} + \frac{[V]}{20} + \frac{[W]}{25} + \frac{[Sn]}{27} + \frac{[Nb]}{43} + \frac{[Ta]}{75} + \frac{[Zr]}{90} - \frac{[Al]}{18}) \text{ (in wt\%) (1)}$$

The minimum value needed to keep the β-phase upon loading has been estimated to be about 3.5 wt% [24], it was therefore concluded that the stress-induced martensitic transformation could be inhibited as a primary deformation mechanism if the $[Fe]_{eq}$ parameter is higher than 3.5 wt%. This equation also shows that iron is the most effective element to increase $[Fe]_{eq}$. Coupling those two design strategies, the composition Ti-7Cr-1Al-1Fe (wt%) was proposed, with a $[Fe]_{eq}$ parameter of 3.53 wt%.

*3.2 Mechanical properties*

Tensile testing (Figure 1) shows that the alloy exhibits an excellent combination of tensile properties with an increased yield strength ($\sigma_{0.2\%}$ = 650 MPa), compared to most of previously reported TRIP/TWIP titanium alloys [1–3,10], a very high ultimate tensile strength (UTS = 1415 MPa), and a remarkably high and stable strain-hardening rate, over the entire range of plastic deformation (Figure 1a). It is worth noting that, for the investigated material, the difference between the ultimate strength and the proof strength UTS- $\sigma_{0.2\%}$ (765 MPa) is higher than $\sigma_{0.2\%}$ itself (650 MPa). This extremely high level of resistance goes along with a high uniform deformation ($\varepsilon_{uniform}$ = 0.38) which is comparable to other TRIP/TWIP titanium alloys [1–3]. The TCAF strain-hardening curve appears to be monotonous, without any "hump", as typically observed in most of TRIP/TWIP assisted titanium alloys (see dash line in Figure 1a). This latter observation is consistent with the noticeable absence of a plateau, in the true stress-strain curve (see solid line in Figure 1a), at the onset of plastic deformation. Since such plateau is generally considered as the signature of the TRIP effect [25], the first assumption is that there is probably no primary TRIP effect in the TCAF alloy upon loading as intended while designing this alloy. Cyclic loading-unloading tests using incremental strain have been performed to further confirm the absence of a significant TRIP effect (Figure 1b). It evidences a hysteresis between unloading and reloading of very small amplitude, quickly reaching a saturation regime at a strain of 0.06 before progressively vanishing. When comparing the amplitude of the hysteresis with the one reported by Brozek in [2] and attributed to the reversion of the stress-induced martensite, it suggests such phenomenon is almost not occurring in the present TCAF alloy. Finally, the elastic part of the loading curve remains almost linear, without double-yielding as sometimes reported, see e.g. [3,26]. All these mechanical evidences strongly suggest that the stress-induced martensitic transformation cannot be considered as a major deformation mechanism in this alloy. This observation is also consistent with the deployed alloy design strategy involving the $[Fe]_{eq}$ parameter.

In the following sections, careful microstructural observations have been performed on deformed TCAF samples to seek for the fundamental mechanisms explaining the remarkable strain-hardening of this alloy that contributes to achieve a large ductility.

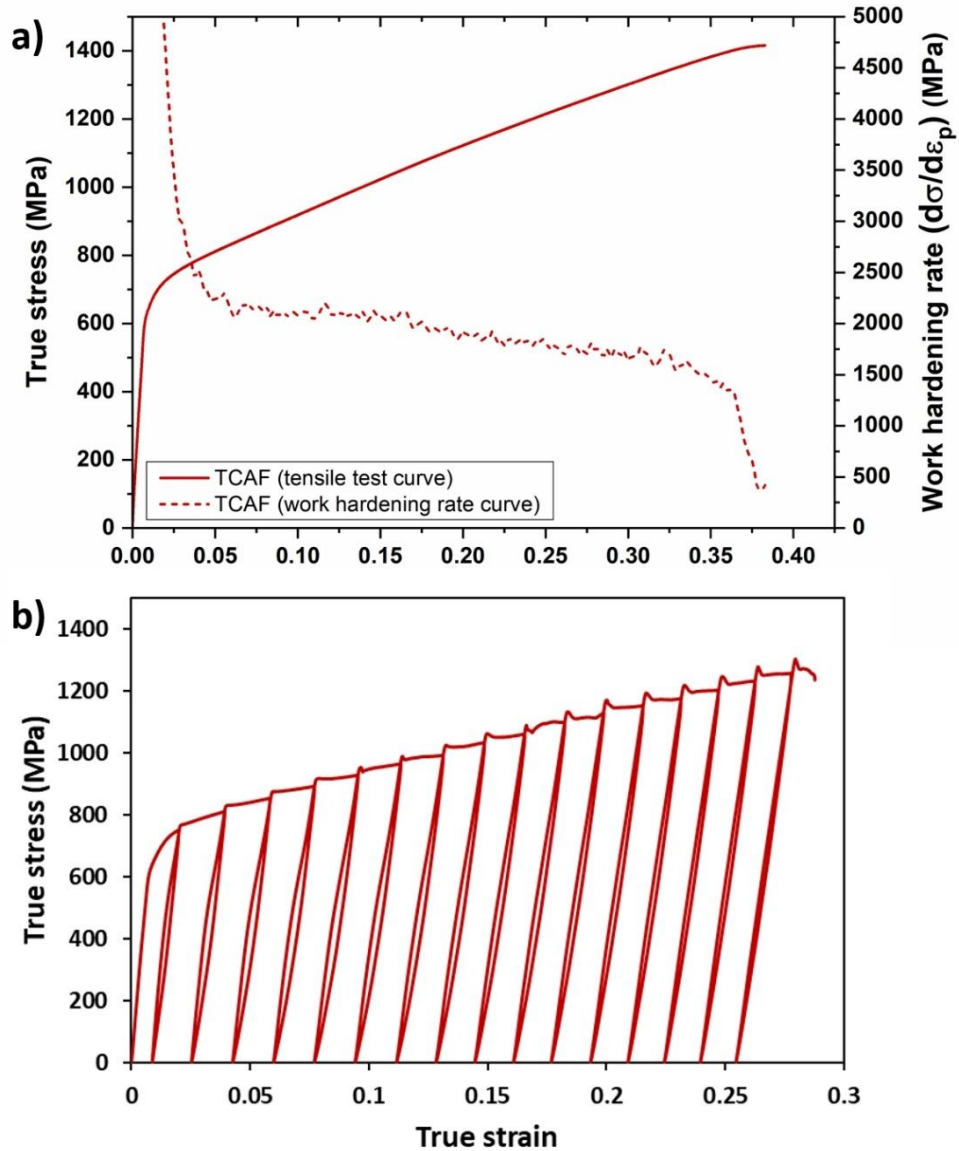

*Figure 1: a) True stress-strain curve (solid line) of the designed alloy (Ti-7Cr-1Al-1Fe) and the corresponding strain-hardening rate curve (dash line). b) Cyclic tensile test with an increment of 2% plastic strain.*

*3.3 Heterogeneous microstructure and localized deformation in loaded TCAF*

The identification and activation of the various deformation mechanisms in the TCAF alloy are performed based on microstructural observations during *in-situ* tensile tests within the SEM, coupled with simultaneous EBSD mapping. Figure 2 shows the evolution of the microstructure from the initial unloaded state to 3% and 4% strain. At 3% strain, {332}<113> twins, which appear to be preferentially localized in certain grains, are observed. Selection of the grains that twin first have been shown to be related to the Schmid factor [27–30], and heterogeneous deformation has previously reported in other TRIP-TWIP Ti-alloys [3]. Nonetheless, some grains present, at 3% strain, a large number of twins (7 twins in Grain 5), and sometimes twins from different twinning systems (3 twinning systems activated in Grain 1), when some neighbors remain undeformed. In agreement with the previously described stress-strain curves (Figure 1), no primary α'' bands (as intragranular precipitation from the β-matrix to accommodate the external stress) have been observed and indexed. This situation is maintained at 4% strain, with some grains that are heavily twinned (see typically grains identified as G1 to G5 in Figure 2b) whereas others display no or very few twins (see for example grains G6 to G9 in Figure 2b).

One can also notice that mechanical twins can be transferred from one grain to another, which has already been studied for {332}<113> twinning in a β-metastable alloy [31] and observed in other alloys with hcp structure [32–34], or stopped at the grain boundary. In good agreement with published results, twin propagation is not observed in the present study for high misorientation angles between adjacent grains (high angle grain boundaries (HAGB), shown in red in Figure 2b), as for instance the twins in G5 that do not transfer to G9 in Figure 2d and f. On the opposite, when the misorientation angle is below 15° (LAGBs shown in blue in Figure 2b), mechanical twins can easily propagate across those boundaries (such as the twin transfer between G3 and G4 in Figure 2d). As the deformation proceeds, twins multiply in the grains, yet the deformation remains heterogeneous. The twin multiplication during the deformation is believed to explain the high and stable strain-hardening rate of the alloy, as proposed by the dynamic Hall-Petch mechanism [35–37] and further investigated in another study [12].

The markedly heterogeneous nature of the deformation mechanisms suggests that grains do not undergo the same local strains for a given macroscopic strain. The heavily twinned grains may undergo a quick local strain-hardening, probably induced by dynamical Hall-Petch effect. In contrast, the neighboring grains without twins display a lower local strain-hardening. The high concentration of twins in some grains, suggesting a high level of deformation, would be expected to lead to quick localization of the deformation and subsequent early fracture of the specimen. However, in the present case, the TCAF alloy exhibits a stable strain-hardening behavior over an extremely large range of plastic deformation.

This raises the question of the response of the material to high local strain fields, leading to its remarkable resistance to fracture. First, their exact location will be investigated, and then the accommodation mechanisms will be unveiled.

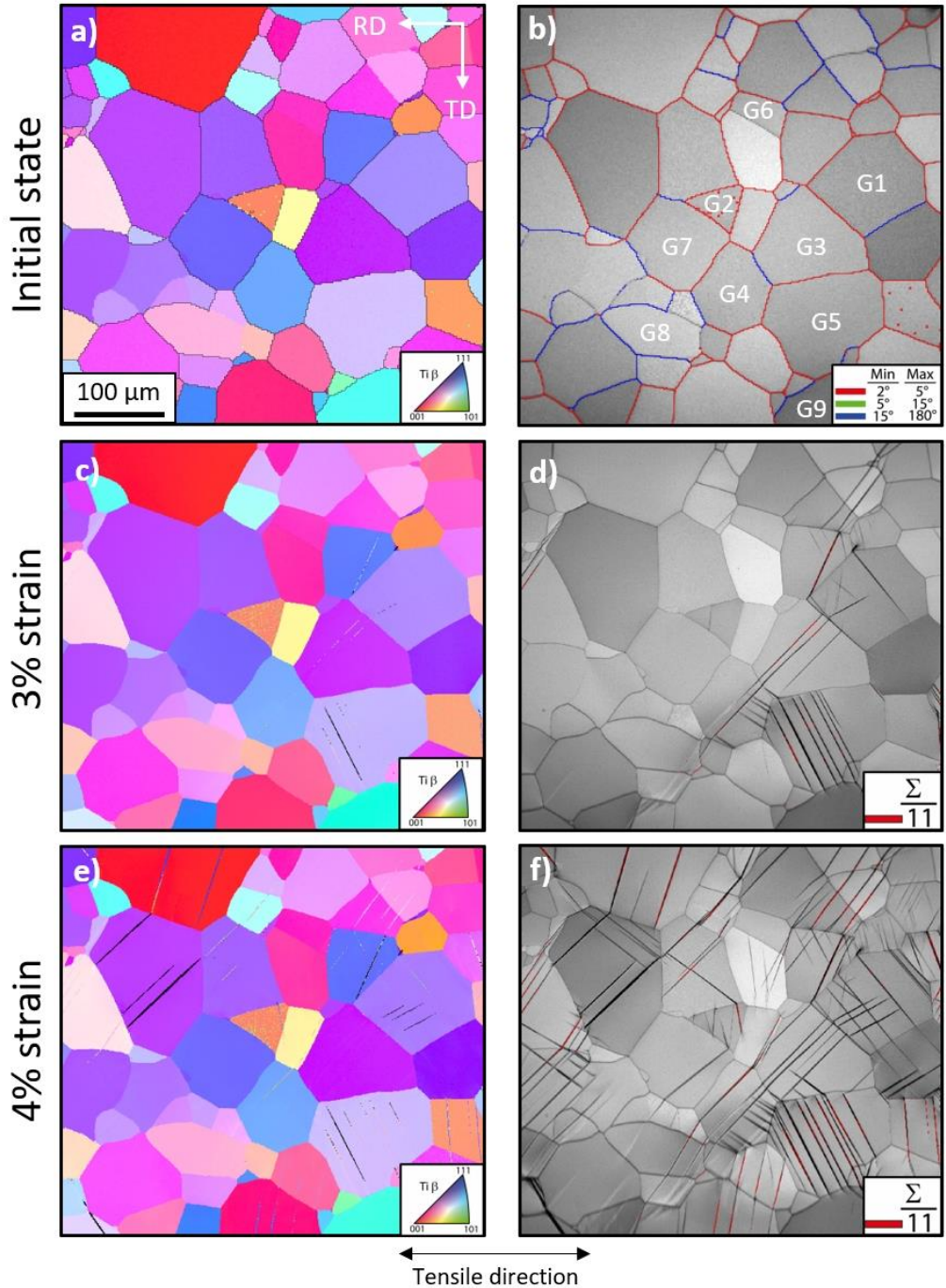

*Figure 2: In-situ observation of the deformation in Ti-7Cr-1Al-1Fe. a) EBSD IPF image along the normal direction at initial stage and b) the corresponding image quality with plotted grain boundaries colored according to the misorientation angle between grains (in blue low angle grain boundaries < 15° and in red high angle grain boundaries > 15°) c) and e) EBSD IPF images along the normal direction at respectively 3% and 4% of macroscopic tensile-strain upon loading and d) and f) the corresponding images qualities. Lines plotted red correspond to 50.5° <110> misorientation (CSL Σ11 equivalent to {332}<113> twins with 4.5° tolerance). (color should be used)*

*3.4 Determination of the location of the highly strained areas*

Location of the high local strain fields in the deformed materials were sought by digital image correlation during *in-situ* tensile test at different deformation stages. This approach allows the microscale strain heterogeneities to be quantified and their distribution within the microstructure to be determined during deformation. This part of the investigation has been dedicated to the early stages of the plastic deformation.

First, EBSD mapping of the initial microstructure of the tested specimen was performed, and provided in Figure 3a and b, allowing to select the region of interest (indicated by a black rectangle). The latter was chosen to select grains with adequate Schmid factor for twinning, as well as grain boundaries. Figure 3c and d show a scan of the final microstructure in the region of interest after 1.98% deformation, after a slight electropolishing of the surface aiming at removing the carbon contaminated layer. The scan of the deformed sample highlights the presence of a dense network of {332}<113> twins in the upper grain.

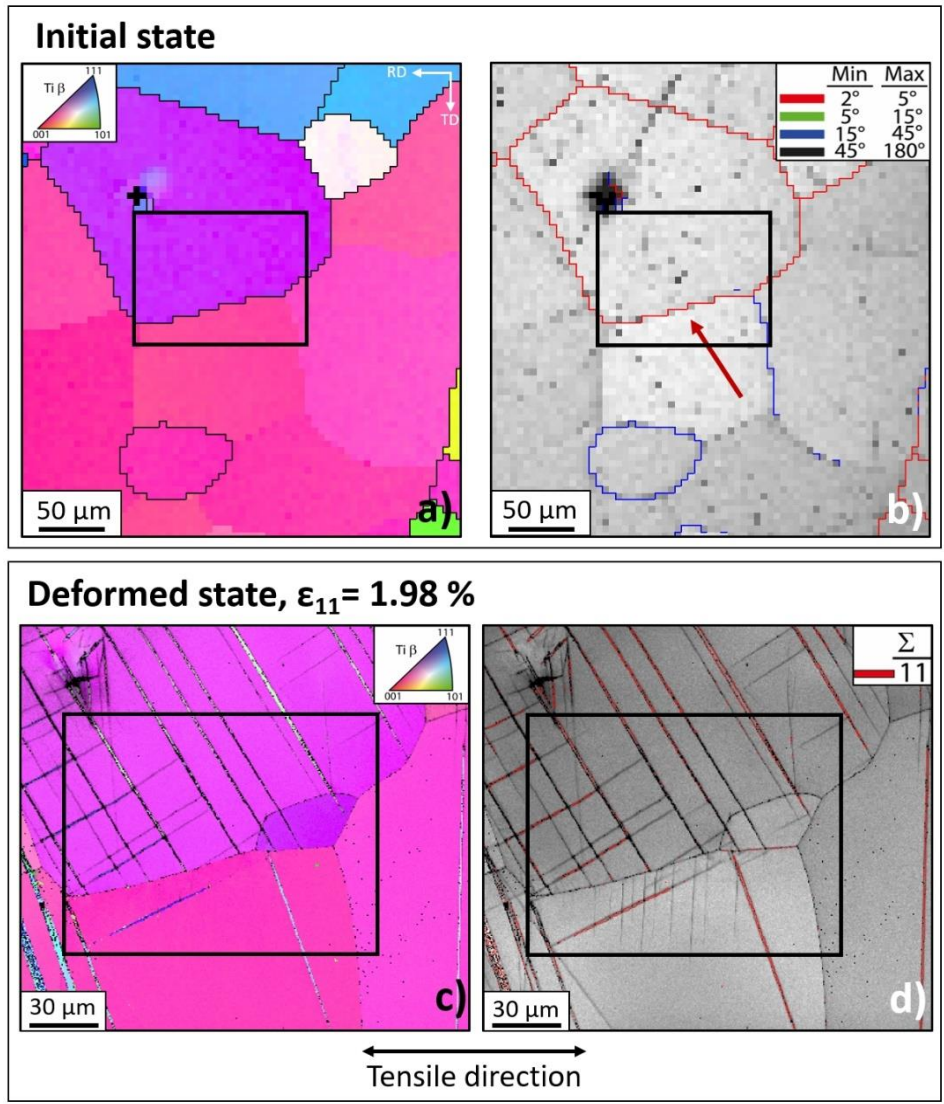

*Figure 3: a) EBSD IPF map along the normal direction at initial stage and b) the corresponding image quality with plotted grain boundaries colored according to the misorientation angle between grains (in blue low angle grain boundaries < 15° and in red high angle grain boundaries > 15°). c) EBSD IPF map along the normal direction after deformation and unloading and d) the corresponding image quality. Lines plotted red correspond to 50.5° <110> misorientation (CSL Σ11 equivalent to {332}<113> twins with 4.5° tolerance). Black squares indicate the reference area for the digital image correlation, see Figure 6.*
*(color should be used)*

Figure 4 details the evolution of the spatial strain distribution within this region of interest during the loading by plotting tensile plastic strain ($\varepsilon_{11}$) maps for two intermediate strains and for the final 1.98% strain of the test. The strain field shows that strain heterogeneities quickly develop at the onset of the plastic deformation (Figure 4a, 0.61% strain). The strain concentration zones, revealed by a color scale, are clearly located at the intersection between twins and in the vicinity of grain boundaries, as well as inside and in the immediate vicinity of {332}<113> twins (Figure 4b and f). This latter observation is probably related to secondary deformation mechanisms, inside primary twins, as previously reported, see e.g. [3,10]. Another striking observation is the strong strain localization found at the intersection of multiple twin variants (see the regions pointed by white arrows in the enlarged views shown in Figure 4f).

Additionally, it is interesting to observe that, by adjusting the range of the (plotted) local tensile plastic strain from [0-0.25] to [0-0.03] in Figure 4d and e, the local strain heterogeneities developing near the grain boundary can be emphasized. Such range also highlights, in the closeups views displayed in Figure 4d and e, how twins are interact with the grain boundary. More interestingly, one can see, in the neighboring grain, a dense network of dislocations revealed by the wavy slip traces observed in the strain map (indicated by a black arrow in Figure 4e). This increased plastic activity on the opposite side of the grain boundary is thought to be caused by the interaction between the twins and the grain boundary. In other words, once the twins hit the grain boundary, dislocations are emitted in the adjacent grains as a result of the local stresses increase.

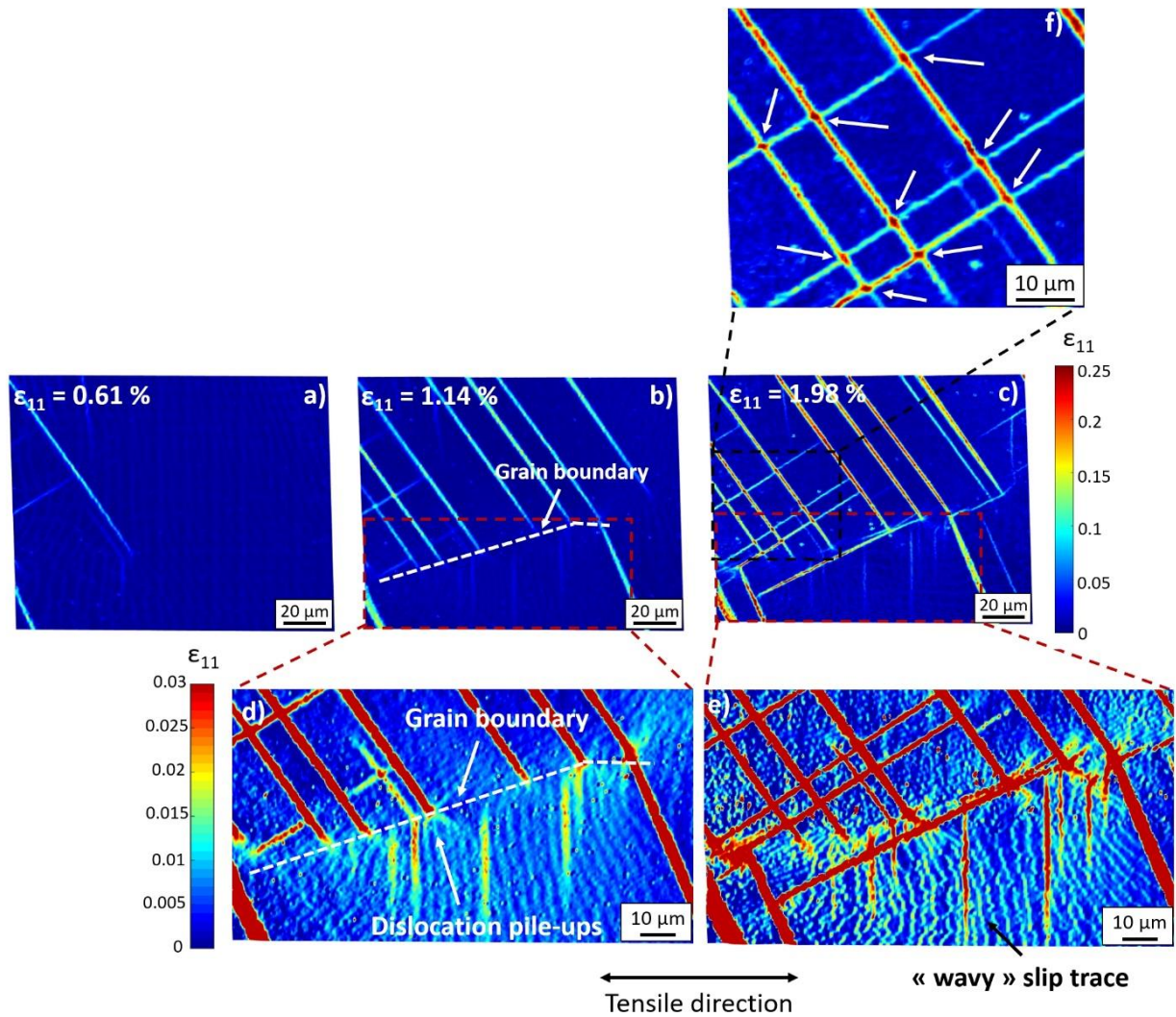

*Figure 4: a) to c) In-situ measurement of the local deformation by digital image correlation at respectively 0.61%, 1.14% and 1.98% of macroscopic tensile plastic strain. d) and e) enlarged views of respectively b) and c) areas in red dash line squares for a reduced scale [0-0.03]. f) enlarged view of c) area in black dash line square. (color should be used)*

Complementary observations are obtained by plotting the GNDs density map obtained by EBSD on a 2% deformed sample (Figure 5). Figure 5a shows 4 grains, separated by high- and low-angle grain-boundaries that deformed by twinning. Transmission of the twin between grain 1 and grain 2, with a minor misorientation, occurs without deviation of the twin (red arrow #1 in Figure 5a). However, grain boundaries between grain 1 and grain 3 and between grain 1 and grain 4 exhibit a high misorientation, approximately equal to 35° and 62° respectively (Figure 5c).

Twins emitted from grain 3, upon hitting the 35° misoriented boundary between grain 1 and grain 3, produce zones of higher GND density in grain 1 (red arrow #2 in Figure 5a). For the grain 4, with a 62° misorientation with grain 1, an orientation gradient is observed in the matrix of grain 1 (Figure 5a), starting where incident twins from grain 2 hit the grain boundary (red arrow #3 in Figure 5a). The IQ map of Figure 5b shows that this zone of grain 1 seems highly affected, which is confirmed by a region dense in dislocation visualized by the GND density map of Figure 5d. After the dislocation-concentrated zone, twins are seen in grain 1 (star signs in Figure Figure 5a). This suggests that the deformation induced by the twins of grain 2 interacting with the grain boundary was so high that dislocations were generated, ultimately triggering the formation of a new twin in grain 1.

Finally, as suggested by the DIC experiment, higher GND concentration is also observed inside the twins and at the intersection between twins (grain 1).

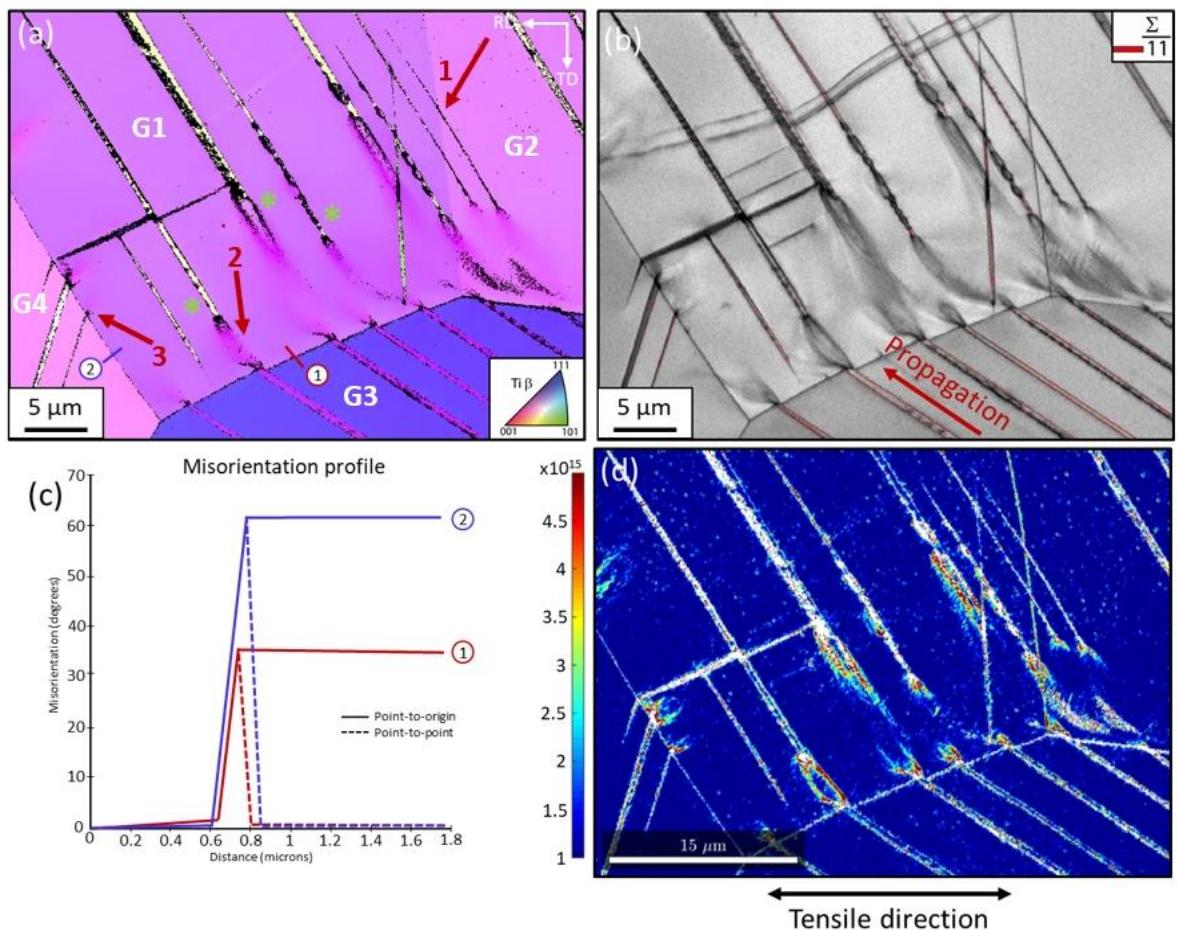

*Figure 5: a) EBSD IPF image along the rolling direction of a 2% deformed Ti-7Cr-1Al-1Fe sample, b) the corresponding image quality where lines plotted red correspond to 50.5° <110> misorientation (CSL Σ11 equivalent to {332}<113> twins with 4.5° tolerance), c) Misorientation profiles along grain boundaries indicated in a) by a red line (point 1) and a blue line (point 2) and d) the GND density map. (color should be used)*

Based on these observations from DIC and dislocation density extracted from EBSD maps, it can be hypothesized that when mechanical twins meet highly misoriented grain boundaries, they induce dislocations in the neighboring grain which will lead to a further increase of the local stresses. If the stress is high enough, then the neighboring grain is able to twin from the head of the stacked dislocations (if the stress-concentration at this head is higher than the critical stress for twin nucleation within the grain), as it can be seen in Figure 4 and Figure 5. Regarding these two figures, it appears that twin propagation is somehow an indirect mechanism involving dislocations piles-up triggering new twins.

Overall, the evolution of the deformed microstructure in the case of the TCAF alloy shows the progressive development of a very dense network of {332}<113> twins as the primary deformation mechanism (Figure 2, Figure 3 and Figure 5). This may explain, as already observed in previous reports [11,12,28,37], the high strain-hardening rate observed in this alloy. DIC and EBSD evidenced that two specific locations thus seem to concentrate most of the higher strains: the intersection between the twins and the grain boundaries, and the intersection between two twins. The interior of the twins will not be discussed in the following, as secondary mechanisms taking place since they have already been investigated extensively [8,23,38–41]. Those two peculiar locations, namely grain boundaries and twin-twin interactions are thus more closely studied in the following, to see how the high strains-fields are accommodated to account for the large ductility of the alloy, and to prevent early fracture.

### *3.5 Microstructural investigation of the high-strained regions*

EBSD was used in a 5% strained specimen to investigate how high-strain regions stand such deformation fields without leading to early fracture.

First, twin intersection was analyzed. EBSD inverse pole figure (IPF) maps of Figure 6a and Figure 6c clearly show intersecting variants of {332}<113> mechanical twins. EBSD indexing becomes challenging at the points of intersection of two twin variants, probably due to the high local strains experimentally evidenced by DIC in the first part (Figure 4f). But mostly, the phase maps of Figure 6b and Figure 6d of this EBSD mapping unveils the presence of a network of α'' needles (in red) that form within the twin (bcc structure, in green) at every twin intersection.

Orientation relationships have been extracted and are consistent with Burgers relationships between α'' and the twinned β-phase.

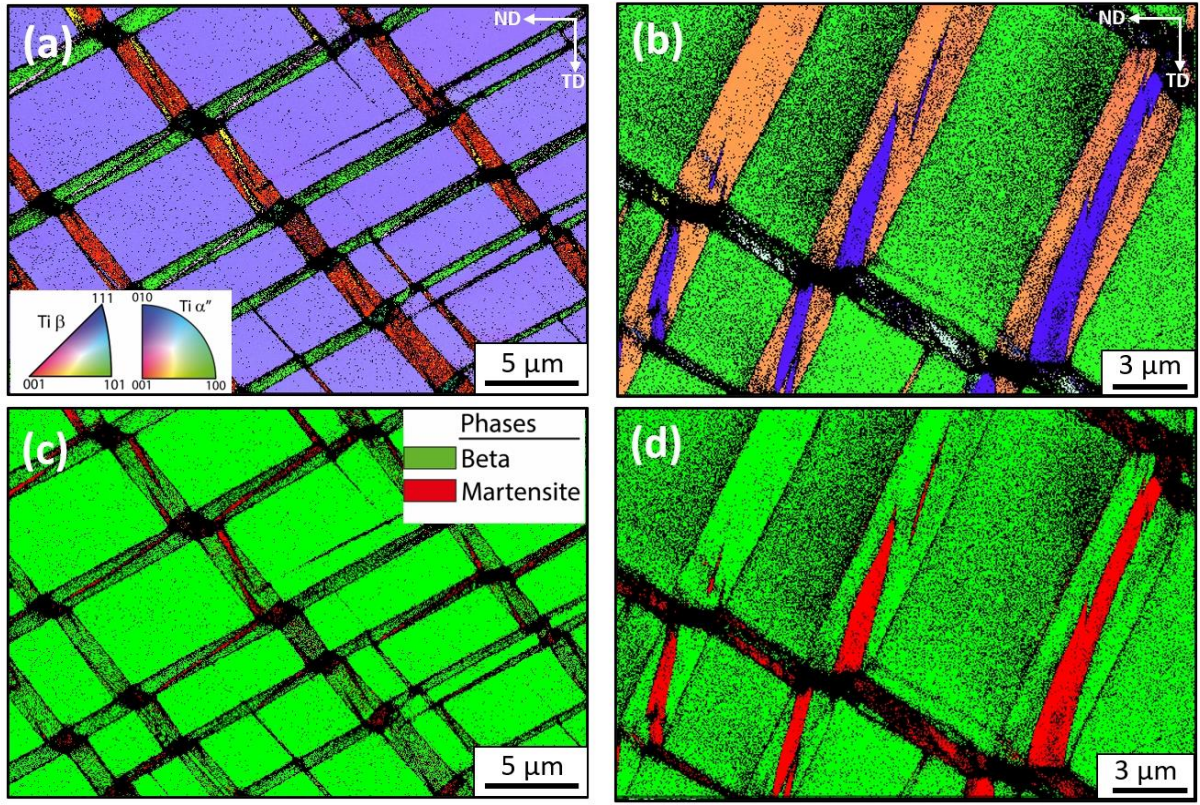

*Figure 6: a) and b) EBSD-IPF images along the rolling direction, corresponding to the tensile direction, of a 5% deformed Ti-7Cr-1Al-1Fe sample and c) and d) the corresponding phase maps with β phase in green and α'' phase in red. (color should be used)*

Then, intersection of twins with a highly misoriented grain boundary (Figure 7a and b) has been mapped, still for a specimen deformed at 5% strain. One can see that twins in the right-side grain cannot cross this highly misoriented grain boundary, as already pointed out and shown in Figure 2, Figure 3 and Figure 5. This induces a high local back-stress and the subsequent generation of a high density of dislocations in the twinned grain, in the vicinity of the boundary interface (Figure 7c). In the neighboring grain, where mechanical twinning is still not triggered at 5% strain, dislocation density is surprisingly low, but the presence of small α'' needles, in red in Figure 7b and highlighted by a white arrow, is evidenced. A reasonable explanation may be that these α'' needles are initiated at the grain boundary and emitted in the β matrix to accommodate the local stress/strain incompatibilities between the highly hardened twinned grain (right side) and the neighboring still "soft" grain, at 5% deformation.

Similarly to the observations of Figure 6, martensite needles are also observed on the phase map of Figure 7c at the intersection of twin boundaries.

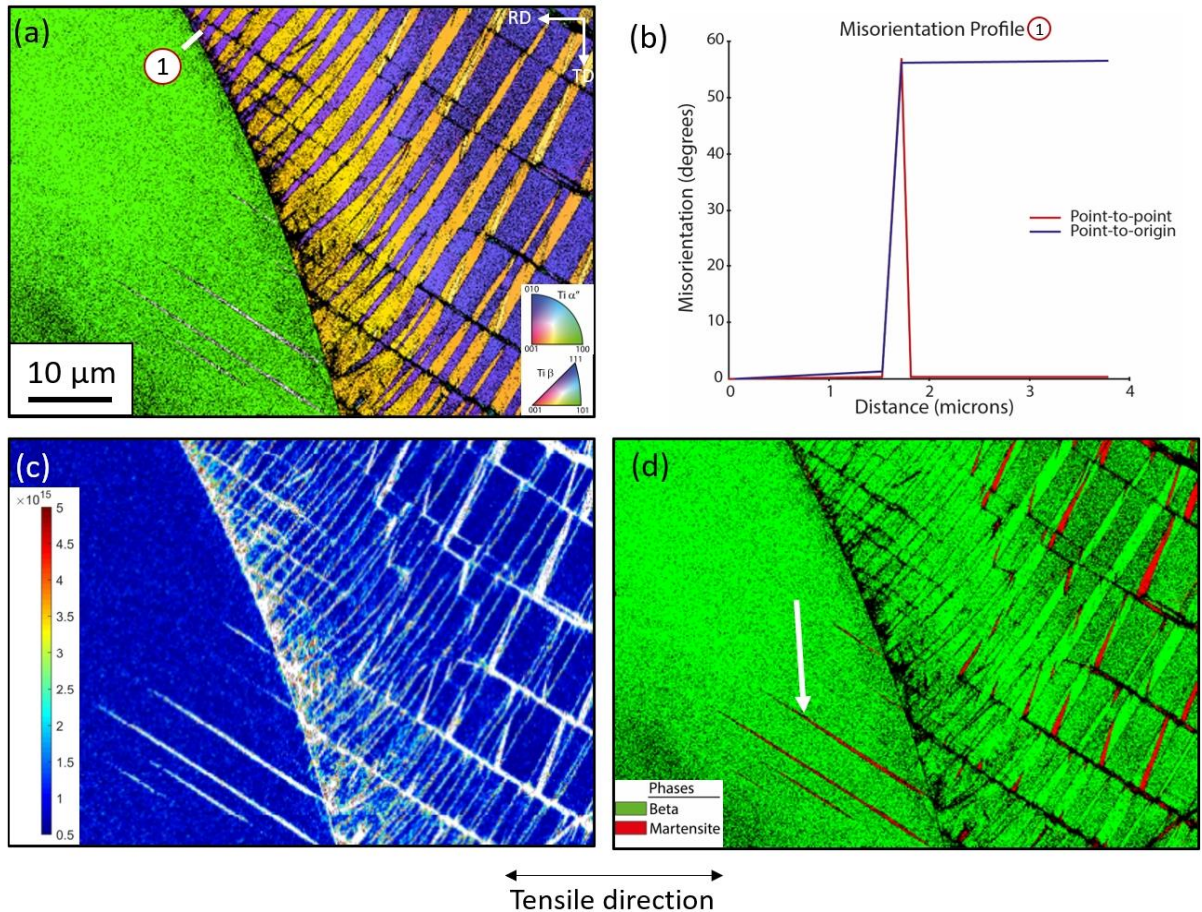

*Figure 7: a) EBSD-IPF images along the rolling direction of 5% deformed Ti-7Cr-1Al-1Fe sample, b) Misorientation profile across the grain boundary indicated by a white line in a). c) Corresponding GND density map. d) Corresponding phase map with β phase in green and α'' phase in red and c) d). (color should be used)*

Figure 6 and Figure 7 show that at the highly constrained zones, identified by DIC as the intersection between deformation twins and intersection between deformation twins and grain boundaries, martensite is formed as thin needles. These observations are in line with observation of martensitic transformation at the interface between α and β of a Ti-10-2-3 alloy [7]. In good agreement with the results of [7], it is suggested that martensite forms in the strained areas to relax the intense internal strain-fields. By bringing an accommodation mechanism, such areas should see the local strains decreasing, which prevents early failure of the material, and rationalizes the large ductility of these alloys. It is interesting to notice that in this material, as intended in our alloy design strategy, martensite is not obtained as a primary deformation mechanism – that accommodates the macroscopic strain – but as a relaxation mechanism. Therefore, the yield strength remains rather high, compared to that of TRIP/TWIP alloys with TRIP as a primary deformation mechanism, and where the formation of martensite occurs at low stress, leading to rather low yield strengths [1,3,10,26,42].

## 4. Conclusions

A new β-metastable titanium alloy has been designed using a combination of the d-electron method and the [Fe]eq parameter with an emphasis on suppressing the TRIP effect and promoting the TWIP effect, with the aim to study understand how TWIP alloys, which display a strong strain localization, can display a large ductility as well as a stable strain-hardening rate over an extremely large deformation range. The results are summarized as follows:

- Tensile tests showed that the alloy displays a yield strength at 650 MPa, over 35% of ductility, and a high and steady strain-hardening rate.
- In-situ EBSD evidenced that the alloy deforms by {332}<113> deformation twinning, with localization of the deformation in some grains, and therefore a heterogeneous distribution of the deformation macroscopically in the alloy. The formation of a dense twinning networks explains the high and stable strain-hardening rate, by dynamic Hall-Petch effect.
- DIC was used to reveal the locations where strain concentrates the most: the intersection between two deformation twins variants and between deformation twins and grain boundaries.
- Detailed EBSD evidenced that martensitic transformation happens as a secondary mechanism at these highly-strained places. This is used to rationalize the extra-large ductility of this material.

This study therefore proposes an explanation for the large ductility of TWIP Ti-alloys: since micro-scale martensitic transformation occurs at places concentrating intense and local strain fields, it is suggested that this phase transformation occurs to relax such constrained locations, hence allowing the deformation to proceed without materials failure. While the fracture is delayed, deformation keeps proceeding in the material by twin multiplication, hence maintaining a high and stable work-hardening. With this study, it is believed that the mechanical behavior of this family of materials can be better understood, leading to a broad range of future optimizations and applications.

**Data availability**

The raw/processed data required to reproduce these findings cannot be shared at this time as the data also forms part of an ongoing study.

**Acknowledgments**

This work was supported by French National Agency for Research (ANR) in the ANR-15-CE08-0013 TITWIP project. YD and FP are grateful to SAFRAN and TIMET companies for their financial support on the YD Ph.D. grant. F. Charlot from the Grenoble-INP characterization platform (CMTC) is gratefully acknowledged for his help during in situ mechanical testing.

**Author contributions**

**Y. Danard**: Writing - original draft, Conceptualization, Investigation. **G. Martin**: Investigation, Formal analysis, Writing - review & editing. **L. Lilensten**: Writing - review & editing, Formal analysis. **F. Sun**: Investigation, Formal analysis. **A. Seret**: Software, Formal analysis. **R. Poulain**: Data curation. **S. Mantri**: Formal analysis. **R. Guillou**: Resources, Formal analysis. **D. Thiaudière**: Resources, Formal analysis. **I. Freiherr von Thüngen** : Supervision. **D. Galy**: Methodology, Formal analysis. **M. Piellard**: Supervision. **N. Bozzolo**: Investigation, Formal analysis. **R. Banerjee**: Formal analysis. **F. Prima**: Conceptualization, Supervision, Writing - review & editing.

**Funding**

This work was supported by the French National Agency for Research (ANR) in the ANR-15-CE08-0013 TITWIP project. YD and FP are grateful to SAFRAN company for its financial support on the YD Ph.D. grant. IF and MP declare that this study received funding from SAFRAN TECH through the YD Ph.D. grant. The funder was not involved in the study design, collection, analysis, interpretation of the data, the writing of this article or the decision to submit it for publication.